\documentclass[aps,prl,twocolumn,showpacs,floatfix,preprintnumbers,amsmath,amssymb,superscriptaddress]{revtex4}
\usepackage{epsfig}
\usepackage{graphicx}
\usepackage{dcolumn}
\usepackage{bm}

\begin{document}
\title{Extreme events on complex networks}
\author{Vimal Kishore}
\email{vimal@prl.res.in}
\affiliation{Physical Research Laboratory, Navrangpura, Ahmedabad 380 009, India.}

\author{M. S. Santhanam}
\email{santh@iiserpune.ac.in}
\affiliation{Indian Institute of Science Education and Research, Pashan Road, Pune 411 021, India.}

\author{R. E. Amritkar}
\email{amritkar@prl.res.in}
\affiliation{Physical Research Laboratory, Navrangpura, Ahmedabad 380 009, India.}

\date{\today}

\begin{abstract}
We study the extreme events taking place on complex networks.
The transport on networks is modelled using random walks and we compute the
probability for the occurance and recurrence of extreme events on the network.
We show that the nodes with
smaller number of links are more prone to extreme events than the ones
with larger number of links. We obtain analytical estimates and verify them
with numerical simulations. They are shown to be robust even when random
walkers follow shortest path on the network.
The results suggest a revision of design principles and can be used as an
input for designing the nodes of a network so as to smoothly handle an extreme event.
\end{abstract}
\maketitle

Extreme events(EE) taking place on the
networks is a fairly common place experience.
Traffic jams in roads and other transportation networks, web servers
not responding due to heavy load of web requests, floods in the network
of rivers, power black outs due to tripping of power grids are some of
the common examples of EE on networks.
Such events can be thought of as an emergent phenomena due to the transport
on the networks.
As EE lead to losses ranging from financial and productivity
to even of life and property \cite{eevent1}, it is important to estimate probabilities for
the occurance of EE and, if possible, incorporate them to design
networks that can handle such EE.

Transport phenomena on the networks have been studied vigorously
in the last several years \cite{trans1,rw1} though they were not focussed on the
analysis of EE. However, one kind of extreme event in the
form of congestion has been widely investigated \cite{congest}. For instance, a typical
approach is to define rules for {\sl (a)} generation and transport of traffic
on the network and {\sl (b)} capacity of the nodes to
service them. Thus, a node will experience congestion when its capacity
to service the incoming 'packets' has been exceeded \cite{cong1}. In this framework, several
results on the stability of networks, cascading failures to congestion
transition etc. have been obtained. Extreme event, on the other hand, is
defined as exceedences above a prescribed quantile and is not necessarily
related to the handling capacity of the node in question. It arises from natural
fluctuations in the traffic passing through a node and not due to constraints imposed
by capacity.  Thus, in rest of
this paper, we discuss transport on the networks and analyse the probabilities
for the occurance of EE arising in them
without having to model the dynamical processes or prescribe capacity at each of the nodes.

The transport model we adopt in this work is the random walk on complex networks \cite{rw1}.
Random walk is of fundamental importance in statistical physics though
in real network settings many variants of random walk could be at work \cite{rws}. For
instance, in the case of road traffic, the flow typically follows a fixed, often
shortest, path from node $A$ to $B$ and can be loosely termed deterministic. As we show in
this paper, thresholds and
corresponding probabilities for the EE depend on such details as the
operating principle of the network. Thus, given the operational principle
of network dynamics, {\it i.e.}, deterministic or probabilistic or a combination
of both, can the nodes of the network be designed to have sufficient capacity
to smoothly handle EE of certain magnitude? We show that we can
obtain {\sl apriori} estimates for the volume of transport on the nodes given the static
parameters and operating principle of the network.
Currently, for univariate time series, there is a widespread
interest on the extreme value statistics and their properties,
in particular in systems that display long memory \cite{mss}.
Thus, we place our results in the context of both the random walks and EE
in a network setting.

We consider
a fully connected, undirected, finite network with $N$ nodes with $E$ edges. The
links are described by an adjacency matrix $\mathbf{A}$ with whose elements $A_{ij}$
are either 1 or 0 depending on whether $i$ and $j$ are connected by a link or not respectively.
On this network,
we have $W$ non-interacting walkers performing the standard random walk. A random
walker at time $t$ sitting on $i$th node with $K_i$ links can choose to hop to any of the
neighbouring nodes with equal probability. Thus, transition probability for
going from $i$th to $j$th node is $A_{ij}/K$. We can write down a master
equation for the $n-$step transition probability of a walker starting from
node $i$ at time $n=0$ to node $j$ at time $n$ as,
\begin{equation}
P_{ij}(n+1) = \sum_k \frac{A_{kj}}{K_k}  ~P_{ik}(n)
\end{equation}
It can be shown that the $n-$step time-evolution operator corresponding
to this transition, acting on an initial distribution, leads to stationary distribution
with eigenvalue unity \cite{rw1} and it turns out to be
\begin{equation}
\lim_{n\to\infty} P_{ij}(n) = p_j = \frac{K_j}{2E},
\label{statdist}
\end{equation}
 The existence of stationary distribution is crucial
for defining EE.
Physically, the time-independent probability in Eq. \ref{statdist} implies that more walkers will
visit a given node if it has more links.

Now we can obtain the distribution of random walkers on a given node. We ask for the
probability $f(w)$ that there are $w$ walkers on a given node having degree $K$. 
Since the random
walkers are independent and non-interacting, the probability of encountering
$w$ walkers at a given node is $p^w$ while rest of $W-w$ walkers are distributed
on all the other nodes. This turns out to be binomial distribution given by
\begin{equation}
f(w) = {W \choose w} ~p^w ~ (1-\bar{p})^{W-w}.
\label{binomial}
\end{equation}
Now, the mean and variance for a given node can be explicitly written down as
\begin{equation}
\langle f \rangle = \frac{W K}{2 E}, \;\;\;\;\;\;\;\;\;\;\;\;\;
\sigma^2 = W \frac{K}{2 E} \left(1-\frac{K}{2 E} \right).
\label{meanvar}
\end{equation}
Quite as expected, the mean and the variance depends on the degree of the node
for fixed $W$ and $E$.
Note that $K/2E << 1$ and the relation between the
mean and the variance for walkers passing through node can be
written as $\sigma \approx \langle f \rangle^{1/2}$. This
reproduces the relation proposed in Ref. \cite{bar1}, later shown to have
limited validity \cite{ym}.

One natural extension of the result in Eq. \ref{binomial}
is to account for fluctuations in the number of walkers. We assume that the
total number of walkers is a random variable uniformly distributed in the interval
$[W-\Delta,W+\Delta]$. Then the probability of finding $w$ walkers becomes
\begin{equation}
f^{\Delta}(w) = \sum_{j=0}^{2 \Delta} \frac{1}{2 \Delta + 1} {\widetilde{W}+j \choose w} ~p^w ~
(1-p)^{\widetilde{W}+j-w},
\end{equation}
where $\widetilde{W}=W-\Delta$. The mean and variance of this distribution can be obtained
as,
\begin{eqnarray}
\langle f^{\Delta} \rangle  & = & \langle f \rangle, \\
\sigma^2_{\Delta} & = & \langle f^{\Delta} \rangle \left[ 1 + \langle f^{\Delta} \rangle \left\{ \frac{\Delta^2}{3W^2} +
            \frac{\Delta}{3W^2} - \frac{1}{W} \right\} \right].
\end{eqnarray}

In the spirit of extreme value statistics, an extreme event is one whose probability of
occurance is small, typically associated with the tail of the probability
distribution function. In the network setting,
we will apply the same principle to each of the nodes.
Based on Eqns \ref{binomial}-\ref{meanvar}, we will
designate an event to be extreme if more than $q$ walkers traverse
a given node at any time instant. Notice that necessarily the cut-off
$q$ will have to depend on the node (or rather, the traffic flowing 
through the node) in question. Applying uniform threshold
independent of the node will lead to some nodes always experiencing an
extreme event while some others never encountering any extreme event at all.
Hence we define the threshold for extreme event to be $q= \langle f \rangle + m \sigma$, where $m$
is any real number. Then, the probability for extreme event can be obtained as
\begin{equation}
F(K) =  \sum_{j=0}^{2 \Delta} \frac{1}{2 \Delta + 1} \sum_{k=\lfloor q \rfloor + 1}^{\widetilde{W}+j}
    {\widetilde{W}+j \choose k} ~p^k ~ (1-p)^{\widetilde{W}+j-k},
\label{exprob1}
\end{equation}
where  $\lfloor u\rfloor$ is the floor function
defined as the largest integer not greater than $u$.

\begin{figure}[t]
\includegraphics*[width=2.1in,angle=-90]{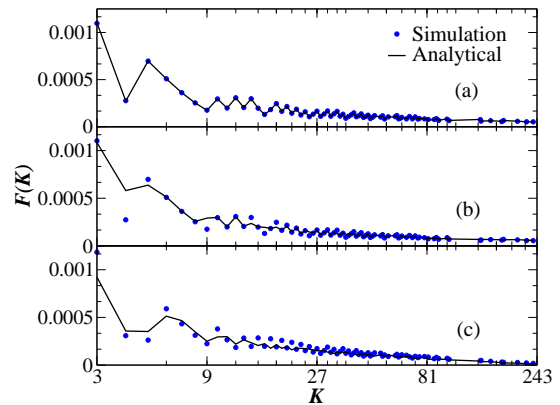}
\caption{Probability for the occurance of extreme events(EE) as a function of degree $K$ of
a node for (a) $\Delta = 0$, (b) $\Delta =0 .01W$ and (c) $\Delta = 0.1W$.  The threshold for EE is
$q= \langle f \rangle +4\sigma$. The solid circles are obtained from
simulations and the solid lines from analytical result in Eq. \ref{exprob1}. All the numerical results 
shown in this paper are obtained with a scale-free
network (degree exponent $\gamma=2.2$) with $N=5000$ nodes, $E=19815$ vertices and $W=2E$ walkers
averaged over 100 realisations. Each realisation corresponds to a new set of randomly
chosen initial conditions to begin the random walk.
}
\label{pex1}
\end{figure}

It does not seem possible to write this summation in closed form.
However, for the special case when $\Delta=0$ Eq. \ref{exprob1} simplifies to
\begin{equation}
F(K)  =  \sum_{k=\lfloor q \rfloor + 1}^W f(K) = I_p \left( \lfloor q \rfloor + 1, w - \lfloor q \rfloor \right)
\label{exprob2}
\end{equation}
where $I_p(.,.)$ is the regularized incomplete Beta function \cite{abs}.
For a given choice of network parameter $E$
and number of walkers $W$, the extreme event probability at any node depends only on its degree.
In Fig \ref{pex1} we show
$F(K)$ as a function of degree $K$ superimposed on the results obtained
from random walk simulations. The agreement between Eq. \ref{pex1} and the simulated results is quite good.
Further, each point in the figure represents an average over all the nodes with the
same degree. We emphasise that the oscillations seen in Fig \ref{pex1} are inherent in the
analytical and numerical results and not due to insufficient ensemble averaging.

\begin{figure}
\includegraphics*[width=2.2in,angle=-90]{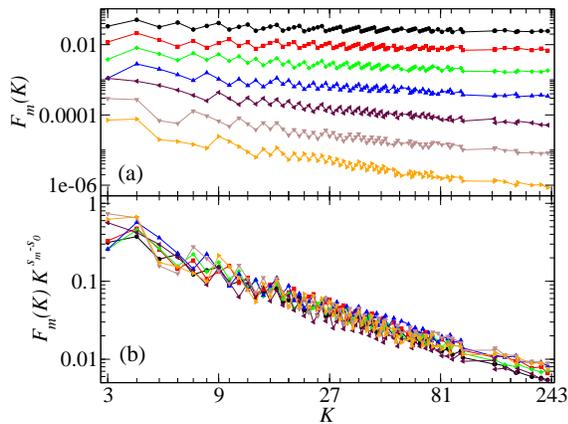}
\caption{(Color online) Probability for extreme events(EE) for several
values of threshold $q=\langle f \rangle + m\sigma$. (a) shows the extreme event
probabilities in log-log plot obtained from simulations with $\Delta=0$. while (b) shows Scaling for the same.
$S_0$ represents the reference slope with $m=2$.
The threshold applied for curves from top to bottom are $m = 2,2.5,3,3.5,4,4.5,5$.}
\label{scale1}
\end{figure}

An important feature of this result is that the nodes with smaller degree ($K < 20$)
reveal, on an average, higher probability for the occurance of EE as
compared to the nodes with higher degree, say, $K > 100$. By careful choice of
parameters, the probability $F(K)$ can differ by as much as an order
of magnitude. This runs contrary
to a naive expectation that higher degree nodes garner
more traffic and hence are more prone to EE. While the former contention is still true in the
random walk model we employ but the results here indicate that the latter one is not generally correct.
As shown in Fig. \ref{pex1}, this feature is robust even when the number of walkers
becomes a fluctuating quantity.
It must be pointed out that Eq. \ref{exprob1}-\ref{exprob2} for the extreme event
probability does not depend on the parameters related to the topology of
the network. Thus, even though the simulation results are shown for scale-free graphs,
it holds good for other types of graphs (not shown here) with random and small world topologies.
However, the difference in probabilty between higher and lower degree nodes
is not pronounced in the case of random graphs.

%In practice, it is difficult to measure flux data at every instant of time on all the
%nodes of a network. In most cases, the cumulative flux for every $M$ time units is measured
%and flow data at a higher temporal resolution is not available.
%Then, the relevant quantity is the probability for EE if the minimal
%resolution of flux data is $M$ time units. In this case too, the probability for EE
%displays a similar feature shown in Fig. \ref{pex1}. The numerical simulation
%results are shown in Fig. {} with $M=10$ and $\delta W=1\% ~\mbox{and} ~ 10\%$.

The threshold $q$ that defines an event to be extreme depends on the traffic
flowing through a given node. The choice $q=\langle f \rangle + m\sigma$ is arbitrary.
Now, we show that the extreme event
probability in Eq \ref{exprob2} scales with the choice of threshold
$q$ or, equivalently, $m$. In the Fig \ref{scale1}(a) we show $F_m(K)$
for various choices of $m$ in log-log scale.
Clearly, as $m$ decreases, ignoring the local fluctuations, the curves tend to
become horizontal. Physically, this can be understood in the
following way; $q \to 0$ implies that the threshold for EE decreases and this leads to larger
number of EE and hence higher probability of occurance. In the limiting case of $q=0$, all the
events would be extreme and we see an equal probability of occurance of EE at all the nodes.
The graph in Fig \ref{scale1}(a) suggests that it might be scaling with respect to $q$ or $m$.
Starting from Eq. \ref{exprob2},
we were not able to determine the scaling analytically. Hence, we empirically show that the following type
of scaling relation holds for the probability of EE,
\begin{equation}
\frac{F_m(K)}{K^{1-S_m}} = \mbox{constant}
\label{sceqn}
\end{equation}
where $F_m(K)$ represents extreme event
probability for threshold value $q$ with parameter $m$.
In this, $S_m$ is the slope of the curves $F_m(K)$ in the Fig. \ref{scale1}(a).
In Fig. \ref{scale1}(b), we show the effect of scaling for several choices of $q$. Using
Eq. \ref{sceqn} on the simulated data for $\Delta=0$, we find that all the curves
for the probability of EE collapse into one curve to a good approximation.

\begin{figure}
\includegraphics*[width=2.2in,angle=-90]{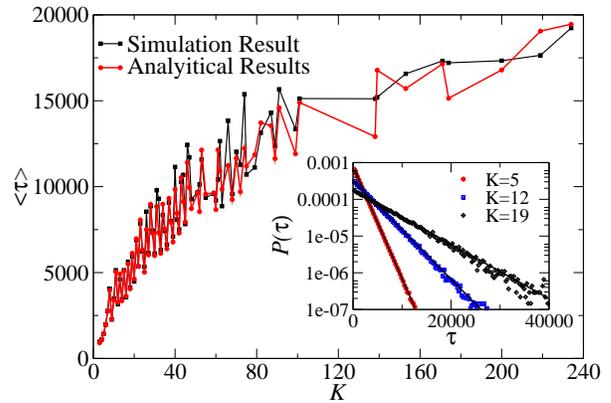}
\caption{(Color online) The inset shows the recurrence time distribution for extreme events
from simulations (symbols) with $\Delta=0$ for nodes with 5, 12 and 19 links.
The solid line is the analytical distribution. The main figure shows the mean
recurrence time as a function of degree $K$.}
\label{recu1}
\end{figure}

In the study of EE, distribution of their return intervals is an
important quantity of interest. This carries the signature of the
temporal correlations among the EE and is useful for hazard
estimation in many areas. We focus on the return intervals for a given node
of the network. Since the random walkers are non-interacting, the events on
the node are uncorrelated. Then, the recurrence time distribution is 
given by $P(\tau) = e^{-\tau/\langle \tau \rangle}$, where the mean recurrence
time is $\langle \tau \rangle = 1/F(K)$. In the inset of Fig. \ref{recu1}, we show
the recurrence time distribution obtained from random walk simulations for three
nodes which have different degrees. In semi-log plot, they reveal an excellent
agreement with the analytical distribution $P(\tau)$ (shown as solid line). The main graph of
Fig. \ref{recu1} shows the mean recurrence time $\langle \tau \rangle$, the only
parameter that characterises the recurrence distribution, as a function
of the degree and it agrees with the analytical result.

\begin{figure}[t]
\includegraphics*[width=2.2in,angle=-90]{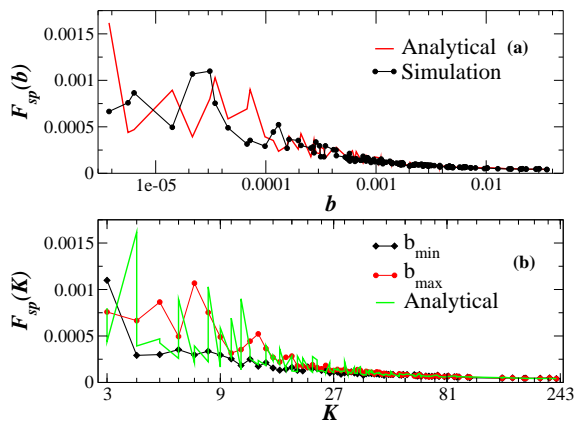}
\caption{Extreme event probability $F_{sp}$ with shortest path
algorithm implemented for random walkers. The data is plotted in two different ways. (a) $F_{sp}(b)$ as a function of betweenness
centrality, (b) $F_{sp}(K)$ as a function of degree $K$ of the node. Nodes with
same value of $K$ can have different betweenness centrality. In (b), in order to
reduce the clutter, for every value of $K$, the extreme event probability for the node with largest (solid circles, red)
and least value (solid square, black) of $b$ is plotted. The fluctuation parameter
$\Delta = 0$.}
\label{spath}
\end{figure}

As pointed out before, many types of flow on the network, such as the information packets
flowing through the network of routers and traffic on roads, use more
intelligent routing algorithms \cite{ral1} rather than performing a random walk.
In order to check the robustness of results in
Eq. \ref{exprob1}-\ref{exprob2}, we implemented the random
walk simulation with the constraint that the traffic from node $i$ to $j$ takes the
shortest path on the network. If multiple shortest paths are available to go
from node $i$ to $j$, the algorithm
chooses any one of them with equal probability. Thus, in this setting, for every
random choice of source-destination pair the paths are
laid out by the algorithm and randomness arises only when multiplicity of shortest
paths are available. In this sense, this can be thought of as walk with a large deterministic
component. We used the shortest path algorithm
developed in Ref. \cite{mej1}. The simulation results, shown in Fig \ref{spath} as solid circles,
are qualitatively similar to the trend displayed in Fig. \ref{pex1}.
In this scenario of predominantly deterministic dynamics due to shortest paths constraint, it is
conceivable that the degree of a node does not
determine the flux passing through it. This role is played by the centrality of the node with respect
to the shortest paths in the network and this is quantified by the betweenness centrality $b$
of a given node \cite{bcent}. Based on this qualitiative argument, the results in
Fig. \ref{spath} can be understood if we replace Eq. \ref{statdist} with $p = \beta b/B$
where $B$ is normalisation factor that depends
on the sum of betweeness centrality of all the nodes on the network. 
From the numerical
simulations, we obtain $\beta \approx 0.94$. Using this $p$ in Eq. \ref{statdist}, we can
go through the same set of arguments as before and obtain $\langle f \rangle, \sigma^2, q$
and the probability for occurance of EE $F_{sp}(b)$ analytically. In Fig \ref{spath}(a) $F_{sp}(b)$ is shown
as solid curve. In Fig \ref{spath}(b) the same data for $F_{sp}(b)$ is shown as a function of $K$ for easier
comparison with Fig \ref{pex1}.
Thus, even with the shortest path algorithm thrown in, the extreme event probabilities are higher
for the nodes with lower degree $(K < 20)$ than for the ones with higher degree $(K > 100)$.

Finally we comment on how these results can be applied as a basis to design
nodes of a network. The central result in this paper in Eq. \ref{exprob1} allows us to
{\sl apriori} estimate the extreme event probabilities. These estimates depend on whether
operating principle of dynamics can be modelled as a purely random walk or
on the basis of shortest paths. If the idea is to avoid congestion or any other
problems arising due to EE of certain magnitude, then these
estimates can be used as an input to the design principles for the nodes.
For instance, for the road traffic that operates broadly on the shortest path principle
the probabilities can be used as an input to design principles (such as higher capacity to nodes)
that will avoid bottlenecks arising from EE of a given magnitude.

In scale-free networks, low degree nodes, which are more prone to experience the EE, form the bulk. 
But design principles and practice generally
focus on the hubs. The results in this work suggest that a revision from such an approach
is necessary. A careful design for the capacity of low degree nodes needs to be given equal importance.
It must be emphasised that incorporating such extreme event estimates in design
principles will only help in better preparedness to meet the expected extreme event.
The extreme events discussed here being due to inherent fluctuations will nevertheless take
place and cannot be avoided.

V. K. thanks IISER, Pune for hospitality during the period when this work was completed.
The numerical simulations reported here were carried on the 3 TEFLOPs and 6 TEFLOPs clusters
 at PRL, Ahmedabad and IISER, Pune respectively.


\begin{thebibliography}{99}

\bibitem{eevent1} S. Albeverio, V. Jentsch and Holger Kantz (Ed.) , {\it Extreme events
in nature and society}, (Springer, 2005).

\bibitem{trans1} S. Boccaletti {\it et. al.}, Phys. Rep. {\bf 424}, 175 (2006);
C. Nicolaides {\it et. al.}, Phys. Rev. E {\bf 82}, 055101(R) (2010);
V. Tejedor,  O. B\'enichou and R. Voituriez, Phys. Rev. E {\bf 80}, 065104(R) (2009); 

\bibitem{rw1}  J. D. Noh and H. Reiger, Phys. Rev. Lett. {\bf 92}, 118701 (2004).

\bibitem{congest} D. De Martino {\it et. al.}, Phys. Rev. E {\bf 79}, 015101(R) (2009);
P. Echenique {\it et. al.}, EPL {\bf 71}, 325 (2005); K. Kim {\it et. al.}, EPL {\bf 86}, 58002 (2009);
B. Tadic {\it et. al.}, EPL {\bf 17}, 2363 (2007); Douglas J. A. {\it et. al.},  Phys. Rev. Lett.
{\bf 94}, 058701 (2005); Liang Zhao {\it et. al.},  Phys. Rev. E {\bf 71}, 026125 (2005);
R. Germano {\it et. al.},  Phys. Rev. E {\bf 74}, 036117 (2006); W. Wang {\it et. al.}, CHAOS {\bf 19},
033106 (2009).

\bibitem{cong1} Wen-Xu Wang et. al., CHAOS {\bf 19}, 033106 (2009)

\bibitem{rws} F. Reif, {\it Statistical Physics}, (McGraw-Hill, 1965) ; C. Song {\it et. al.},
Nature Physics {\bf 6}, 818 (2010); D. Brockmann {\it et. al.}, Nature {\bf 439}, 462 (2006).

\bibitem{mss} J. F. Eichner {\it et. al.}, Phys. Rev. E {\bf 75}, 011128 (2007); M. S. Santhanam
and Holger Kantz, Phys. Rev. E {\bf 78}, 051113 (2008).

\bibitem{mej1} Cormen, T. H.; Leiserson, C. E.; Rivest, R. L.; Stein, Clifford, {\it Introduction to Algorithms} ( MIT Press and McGraw-Hill, 2001) 

\bibitem{bar1}  M. Argollo de Menezes and A. L. Barabási, Phys. Rev. Lett. {\bf 92}, 028701 (2004)
\bibitem{ym} S. Meloni et. al., Phys. Rev. Lett. {\bf 100}, 208701 (2008).

\bibitem{abs} M. Abramowitz and I. A. Stegun, {\it Handbook of Mathematical Functions} (Dover, New York, 1964).

\bibitem{ral1} D. Methi and K. Ramasamy, {\it Network Routing : Algorithms, Protocols and Architectures},
(Morgan Kaufmann, 2007).

\bibitem{bcent} M. Barthelemy, Eur. Phys. J. B {\bf 38}, 163 (2004); H. Wang {\it et. al.},
Phys. Rev. E {\bf 77}, 046105 (2008); S. Dolev {\it et. al.}, J. of ACM {\bf 57}, 25 (2010).

\end{thebibliography}
\end{document}